\documentclass[english,pra,twocolumn, eqsecnum]{revtex4-1}
\usepackage[T1]{fontenc}
\usepackage[latin9]{inputenc}
\usepackage{amsmath}
\usepackage{amssymb}
\usepackage{graphicx}
\usepackage{esint}

\makeatletter

\providecommand{\tabularnewline}{\\}

\@ifundefined{textcolor}{}
{%
 \definecolor{BLACK}{gray}{0}
 \definecolor{WHITE}{gray}{1}
 \definecolor{RED}{rgb}{1,0,0}
 \definecolor{GREEN}{rgb}{0,1,0}
 \definecolor{BLUE}{rgb}{0,0,1}
 \definecolor{CYAN}{cmyk}{1,0,0,0}
 \definecolor{MAGENTA}{cmyk}{0,1,0,0}
 \definecolor{YELLOW}{cmyk}{0,0,1,0}
 }

\makeatother

\usepackage{babel}
\begin{document}

\title{Confinement induced resonances in anharmonic waveguides }

\author{Shi-Guo Peng$^{1,2}$, Hui Hu$^{2}$, Xia-Ji Liu$^{2}$, and Peter
D. Drummond$^{2}$}

\affiliation{$^{1}$Department of Physics, Tsinghua University, Beijing, 100084,
China}

\affiliation{$^{2}$Centre for Atom Optics and Ultrafast Spectroscopy, Swinburne
University of Technology, Melbourne 3122, Australia}
\begin{abstract}
We develop the theory of anharmonic confinement-induced resonances
(ACIR). These are caused by anharmonic excitation of the transverse
motion of the center of mass (COM) of two bound atoms in a waveguide.
As the transverse confinement becomes anisotropic, we find that the
COM resonant solutions split for a quasi-1D system, in agreement with
recent experiments. This is not found in harmonic confinement theories.
A new resonance appears for repulsive couplings ($a_{3D}>0$) for
a quasi-2D system, which is also not seen with harmonic confinement.
After inclusion of anharmonic energy corrections within perturbation
theory, we find that these ACIR resonances agree extremely well with
anomalous 1D and 2D confinement induced resonance positions observed
in recent experiments. Multiple even and odd order transverse ACIR
resonances are identified in experimental data, including up to $N=4$
transverse COM quantum numbers.
\end{abstract}
\maketitle

\section{introduction}

Ultra-cold low-dimensional atomic gases show unique quantum properties,
and have attracted a great deal of interest. For a one-dimensional
(1D) Bose gas\cite{Girardeau1960,LiebLiniger1963,Yang1969}, the finite-temperature
correlations predicted for a Tonks-Girardeau gas\cite{Gangardt,1DCorrelations}
have been experimentally verified\cite{Tolra-NIST-exp,Weiss-exp},
and a cross-over to a non-equilibrium super Tonks-Girardeau gas has
been realized \cite{Haller2009R}. For a two-dimensional (2D) geometry,
the Berezinskii-Kosterlitz-Thouless phase transition was predicted
\cite{Berezins1971D,Kosterlitz1973O}, and subsequently observed in
experiment \cite{Hadzibabic2006B}. In these experiments, one or two
spatial degrees of freedom are removed by introducing tight confinement
via an optical lattice or a tightly focused anisotropic dipole trap. 

Atomic interactions can also be tuned precisely by means of a molecular
Feshbach resonance in an external magnetic field \cite{Feshbach1962A,Kohler2006P,Bloch2008M}.
This allows an effective contact interaction with scattering length
$a_{3D}$ to be created, with scattering lengths that can be varied
over a wide range of positive and negative values. Owing to these
methods, low-dimensional atomic gases now provide a high degree of
control for tests of fundamental many-body physics in reduced dimensions.
These systems are often much simpler than condensed matter physics
experiments, which have complex crystal structure, interactions and
disorder.

Confinement-induced resonances (CIR) are one of the most intriguing
phenomena found in low-dimensional systems. These were first predicted
theoretically by Olshanii \cite{Olshanii1998A}, who considered a
two-body S-wave scattering problem in a quasi-1D trap with cylindrically
symmetric transverse harmonic confinement. The CIR can be understood
as a novel type of Feshbach resonance, where the transverse ground
mode and the manifold of molecular internally excited modes play the
roles of the open and closed channels respectively \cite{Bergeman2003A}.
Related effects occur in mixed dimensional traps \cite{Nishida,Lamporesi}.
A direct generalization of Olshanii's theory to anisotropic transverse
confinement shows that there is only one harmonic CIR (HCIR) resonance,
no matter how large the transverse anisotropy \cite{Peng2010C}. For
large anisotropy, this theory crosses over smoothly to the case of
a quasi-2D trap, where a single HCIR occurs with a negative S-wave
scattering length, $a_{3D}<0$ \cite{Petrov2000B,Petrov2001I,Naidon2007E}. 

There have been a number of related experimental investigations, which
in some cases appear to contradict each other. In the recent Innsbruck
$Cs$ experiment with a quasi-1D geometry \cite{Haller2010C}, two
or more resonances were observed as the transverse confinement became
more and more anisotropic. For a quasi-2D geometry, some experiments
have observed 2D resonances on the attractive side with $a_{3D}<0$
\cite{Frohlich2010R}, while others have resonances on the repulsive
side with $a_{3D}>0$ \cite{Haller2010C,Vale2011}. Both the observation
of multiple resonances and 2D resonances with repulsive interactions
are in disagreement with standard HCIR predictions \cite{Petrov2000B,Petrov2001I,Naidon2007E}. 

In this paper, a detailed explanation is proposed for these anomalous
resonances. The mechanism is that the new resonances are due to center-of-mass
(COM) excitations of molecules or atom pairs. These have a different
character to the excitation of internal molecular degrees of freedom
found in the Olshanii approach and its generalizations. The new COM
resonances can only become coupled to the input state by anharmonic
terms in the trapping potential. Hence, we term these effects anharmonic
confinement induced resonances (ACIR). The ACIR resonances cannot
occur in harmonic traps due to Kohn's theorem \cite{Kohn}. However,
they are certainly observable in current ultra-cold atomic physics
experiments, which have relatively large anharmonicities.

The coupling of the COM motion to the relative motion gives additional
degrees of freedom not found with parabolic traps. This causes a series
of additional scattering resonances due to the mixing of COM and relative
motion. The nonlinear mixing caused by anharmonic terms in the potential
makes these phenomena analogous to frequency-mixing effects found
in nonlinear optics. They provide a fundamentally new pairing mechanism,
which may lead to new opportunities for quantum engineering in atomic,
photonic or acoustic waveguides. Our results are therefore qualitatively
different to harmonic CIR. We predict both multiple 1D resonances
and 2D resonances with repulsive interactions. Both results are in
quantitative agreement with experiment.

We note that this possibility was also envisaged in three earlier
papers. In Peano et al \cite{Peano}, the general idea is addressed,
although using a different technique, and for a different type of
trap. Kestner and Duan \cite{Kestner2010A} treat anharmonic resonance
in a double well. In a more recent investigation, parallel to our
own, Sala et al \cite{Saenz} have also concluded that the recent
Innsbruck experiments provide evidence for ACIR resonances. The main
differences in the treatment are that we have accurately calculated
the size of the anharmonic resonance shifts, as well as giving quantitative
estimates of relevant parameters. We also compare our theory with
the observed multiple resonances, including even and odd order COM
resonances.

The paper is arranged as follows. We first analyze the types of transverse
excitations available, and the operational processes that can lead
to the observed resonances (Sec.\ref{sec:Two-body-CIR}). In Sec.\ref{sec:Nonlinear-CIR},
a Hamiltonian model of nonlinear CIR is presented, by introducing
anharmonic perturbations in the Hamiltonian. In Sec.\ref{sec:Nonlinear-CIR-1D},
this is analyzed using perturbation theory for the 1D case. The results
are compared to experiments on anisotropic traps, showing the observed
splitting is well-explained with the anharmonic COM resonance ACIR
model. Next, we consider results for the case of large trap anisotropies,
and demonstrate that the observed resonances can be quantitatively
explained with excellent accuracy by considering multiple resonances
with both even and odd order COM transverse quantum numbers. A similar
calculation is carried out for a quasi-2D system in Sec. \ref{sec:Nonlinear-CIR-2D},
which is also compared with experiment. The main results are summarized
in Sec.\ref{sec:conclusion}.

\section{Two-body CIR physics\label{sec:Two-body-CIR}}

In recent one and two dimensional confinement induced resonance experiments,
there are many observed resonances not explained by conventional CIR
theory. In two dimensions, resonances are observed for $a_{3D}>0$,
(the BEC side of the resonance) which is the opposite to that expected
in the usual theory. Similarly, unexplained multiple resonances occur
for one-dimensional CIR with anisotropic transverse confinement. These
are also not predicted by the simple two-particle  model \cite{Olshanii1998A}
with linear confinement. However, the experiments have some features
not included in this idealized model, and the obvious question is:
which experimental properties are responsible for the additional observed
resonances?

\subsection{Harmonic CIR solutions}

\begin{figure}
\includegraphics[width=0.9\columnwidth]{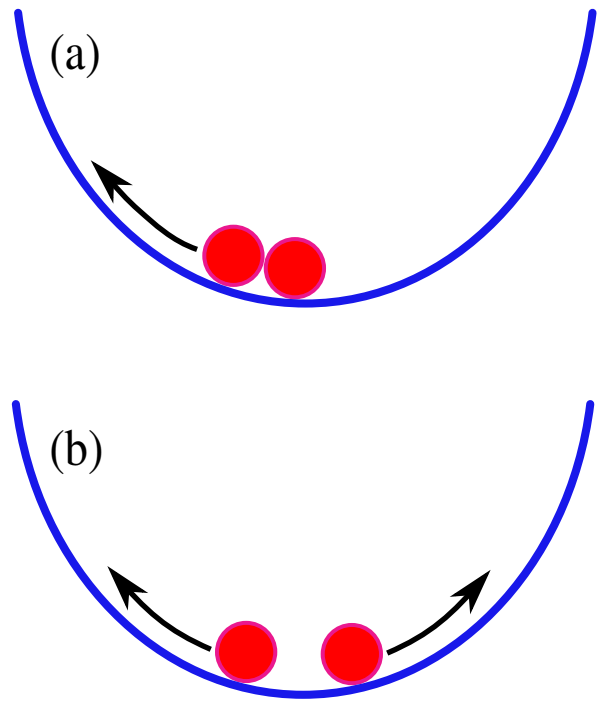}\caption{(Color online) (a) Illustration of an ACIR resonance, in which two
atoms are driven in a molecule-like excited state with a transverse
COM motion in the trap. Both atoms move together, to give a transverse
resonance of the COM degree of freedom. This is only coupled to the
atomic ground state if the potential is anharmonic. (b) Illustration
of an HCIR resonance, in which the atoms move in relative motion,
to give an excited state rather like a conventional vibrational excited
state of a molecule. \label{fig:(Color-online)-Schematic diagram}}

\label{Flo:Innsbruck}
\end{figure}

The possible modes of excitation of a pair of atoms in a transverse
potential are illustrated in Fig.\ref{fig:(Color-online)-Schematic diagram}.
The COM mode is shown in (a), with two atoms moving together. This
is not coupled to the atomic ground state in a harmonic trap, due
to Kohn's theorem. The relative motion mode is shown in (b), with
an excitation of the relative coordinate. This is the usual harmonic
confinement induced resonance in a one dimensional waveguide with
a transverse parabolic potential. The HCIR resonance is simply the
first \emph{internally} excited resonance of the two-body ground state. 

However, there is a subtlety here. Both types of excitation are an
adiabatic continuation of the transversely excited free-particle states,
as the inter-particle interaction is increased. The free-particle
states have both center-of-mass (COM) quantum numbers $N_{x},N_{y},$
and internal quantum numbers, $n_{x},n_{y}$. Therefore, there are
a large number of possible excited states that could, in principle,
be coupled to the atomic ground state in a Feshbach type of resonance.

In a rotationally symmetric model, both the internal and COM excited
states near $2\hbar\omega$ are degenerate in energy, and cannot lead
to distinct resonances even if coupled to the incoming states. If
there is no rotational symmetry, this degeneracy is broken, leading
to the possibility of multiple resonances. Experimentally there is
more than one resonance in the recent Innsbruck experiment \cite{Haller2010C}.
Due to the similarity with the  HCIR predictions \cite{Olshanii1998A},
these were initially identified as conventional harmonic CIR resonances,
corresponding to excitations of internal degrees of freedom.

However, an analysis of the anisotropic case \cite{Peng2010C} indicates
that internally excited two-body states only give rise to only a \emph{single}
CIR. We note that Kohn's theorem \cite{Kohn} prohibits the excitation
of COM resonances from incoming states with no transverse momentum,
if the confinement is parabolic. One may ask why there are not multiple
CIR eigenstates with different internal energies due to the different
confinement strength in orthogonal trap directions. The reason for
this is due to the singular nature of the interaction. 

Consider what happens to a single particle in a rotationally symmetric
potential, which corresponds to the internal quantum numbers in a
relative coordinate picture of a two-particle problem. In a 2D system
of noninteracting particles in a harmonic oscillator potential, the
ground state has $n_{x}=n_{y}=0$. The internally excited states can
therefore be labelled either by their internal harmonic oscillator
quantum numbers for relative motion, or by their angular momentum
quantum numbers. This, one can have $n_{x}=n_{y}=1$, or else a radial
quantum number $n_{r}$ and a magnetic quantum number $m$. Due to
the singular potential at the origin, only S-wave incoming states
with $m=0$ experience any coupling. This leads to a relatively low-lying
excited state due to the coupling, and hence to a single CIR. This
state is adiabatically deformed when the symmetry is broken, without
leading to a second CIR.

This single degenerate CIR exists on the positive side of the Feshbach
resonance i.e, $a_{3D}>0$ for a quasi-1D system. It transfers to
the negative side, i.e., $a_{3D}<0$ for a large asymmetry or a quasi-2D
system. This last conclusion is compatible with other calculations
of 2D CIR. However, these conclusions only take into account the \emph{internal}
energy of a two-particle state, not the COM energy.

\subsection{Anomalous CIR experiments}

In recent bosonic experiments on ultracold $^{137}Cs$ at Innsbruck\cite{Haller2010C},
a strong, transversely anisotropic quasi-1D confinement is used to
create an initially strongly repulsive ($a_{3D}>0$) Tonks gas. This
is followed by a sudden change in $B$-field to a new value, resulting
in a molecular loss signature for a resonance which is confinement
dependent. Numerous multiple confinement-induced resonances are observed.
There is even an unexpected resonance for $a_{3D}>0$ in the two-dimensional
(2D) limit, which has also been measured using release energy data
in fermionic $^{6}Li$ experiments \cite{Vale2011}. All these observations
contradict the harmonic waveguide theory given above. 

However, it is important to recognize that the waveguide potential
in these experiments are generally anharmonic, so that Kohn's theorem
does not apply. Hence, there are more degrees of freedom available
for excitation, since the COM quantum numbers must now be included
in the description. 

The interesting issue is whether these observed CIR effects can be
explained as anharmonic (ACIR) resonances due to center-of-mass excitations
of resonant bound states\cite{Peano,Kestner2010A,Saenz}. This is
illustrated in Fig.\ref{fig:(Color-online)-Schematic diagram}, by
the two atoms moving together in (a). Such effects can only occur
in an anharmonic trapping environment, which allows coupling between
incoming scattering states with zero transverse excitation and an
outgoing transverse COM excitation.

In another 2D experiment on ultracold $^{40}K$ at Cambridge\cite{Frohlich2010R},
a CIR resonance occurs on the attractive side of a Feshbach resonance,
as expected. This experiment has a much lower anharmonicity than the
Innsbruck experiment. It also uses a different technique to identify
the resonance, employing RF spectroscopy rather than molecular losses.
Thus, there are distinct resonance signatures used in the two published
experiments. The Cambridge data appears to show evidence for anomalous
resonance features on the $a_{3D}>0$ side of the Feshbach resonance,
but this effect is greatly reduced compared to the Innsbruck observations. 

All Feshbach bound states have a bound molecular fraction\cite{Drummond}
in which the atoms have a small separation. We conjecture that this
molecular fraction is larger for COM ACIR resonances (a), where the
atoms are in a relative ground state - compared to internally excited
HCIR resonances (b), where the atoms are in a relative excited state.
This would mean that COM excitations would have a relatively larger
3-body recombination loss due to molecule formation. This is precisely
the signature of the resonances used in the Innsbruck experiments.
RF spectroscopy, used in the Cambridge experiments, on the other hand
has different characteristics. Thus, it is not unreasonable to expect
the two experiments to have a different relative sensitivity to internal
and COM molecular resonances.

\section{Anharmonic Waveguides \label{sec:Nonlinear-CIR}}

For technical reasons explained below, current CIR experiments typically
involve an anharmonic confinement mechanism. In such cases, the excitation
of a COM degree of freedom can be coupled to input states with no
transverse COM excitation. This coupling mechanism would explain observed
anomalies such as multiple resonances, different to those predicted
using the standard parabolic confinement theory. 

There are three possibilities which might allow resonant coupling
to additional confinement induced bound-states through nonlinear mechanisms:
\begin{enumerate}
\item The Kohn theorem only applies for parabolic confinement. Experimentally
the optical confinement is sinusoidal and/or Gaussian, not parabolic.
This allows direct coupling to $N_{x}=2$ states, or even $N_{x}=1$
states, depending on the type of anharmonicity.
\item In some experiments the width of the Feshbach resonance is comparable
to the separation of the transverse modes, allowing contributions
from the molecular bound state channel as well as the atomic channel.
This still requires anharmonic coupling to access bound states having
COM transverse energy.
\item At high density the mean-field background potentials of the other
atoms may provide an anharmonic effective potential which is not parabolic.
In such cases, one may expect the collective oscillation frequencies
to play a role.
\end{enumerate}
Coupling to COM excitations is possible whenever the transverse response
is nonlinear, and the potential departs from a parabolic shape. This
is not inconsistent with the ultra-cold atomic physics experiments,
which generally involve sinusoidal laser trapping potentials. These
are only approximately parabolic in the strongly confined limit. It
is this possibility of nonlinear CIR due to anharmonic confinement
which is explained below. Although related theoretical work that has
been carried out includes one or the other of these effects, it is
important to include \emph{both }anisotropy and anharmonicity to fully
explain the observed ACIR resonances.

\subsection{Hamiltonian}

In this paper, we consider the simplest model of nonlinear CIR, with
a single-channel S-wave interatomic potential and an anharmonic trapping
potential. We do not take into account either many-body corrections
or explicit molecule formation channels\cite{Drummond}. This model
is therefore applicable to relatively dilute quantum gases with a
broad Feshbach resonance. To model the nonlinear CIR effect in greater
detail, consider two atoms with mass $m$ which are anisotropically
confined in the transverse direction, and can almost freely move in
the $z$ direction. Such a model can treat both a quasi-1D and quasi-2D
experiment, by taking one of the trapping frequencies to zero. The
Hamiltonian of the two atoms in a quasi-1D system is, therefore:
\begin{equation}
H=H_{1}+H_{2}+U\left(\mathbf{r}_{1}-\mathbf{r}_{2}\right)\,,\label{eq:1D 1}
\end{equation}
where 
\begin{equation}
U\left(\mathbf{r}\right)=g_{3D}\delta(\mathbf{r})\frac{\partial}{\partial r}r
\end{equation}
 is the atomic interaction for S-wave scattering described by a zero-range
pseudo-potential with interaction strength $g_{3D}=4\pi\hbar^{2}a_{3D}/m$
corresponding to a scattering length of $a_{3D}$, and $H_{i}$ is
a single-particle Hamiltonian including external potential and kinetic
energy terms.

\subsubsection{Anharmonic confinement potential}

For dipole trapping experiments with 1D and 2D optical lattices, the
trapping potential near a potential minimum at $\mathbf{r}=0$ is
due to an optical standing wave of form:
\begin{eqnarray}
U^{ext}\left(\mathbf{r}\right) & = & V_{x}\left(\mathbf{r}\right)\sin^{2}\left(\frac{2\pi x}{\lambda_{x}}\right)+V_{y}\left(\mathbf{r}\right)\sin^{2}\left(\frac{2\pi y}{\lambda_{y}}\right)\nonumber \\
 & \approx & \frac{1}{2}m\left[\omega_{x}^{2}x^{2}\left(1+\frac{\alpha_{x}x^{2}}{d_{x}^{2}}+\frac{\mathbf{r}\cdot\nabla V_{x}}{V_{x}^{0}}+\ldots\right)\right.\nonumber \\
 &  & \left.+\omega_{y}^{2}y^{2}\left(1+\frac{\alpha_{y}y^{2}}{d_{y}^{2}}+\frac{\mathbf{r}\cdot\nabla V_{y}}{V_{y}^{0}}+\right)\right]
\end{eqnarray}

Here, $V_{x,y}\left(\mathbf{r}\right)$ are the two orthogonal slowly
varying potential energy envelopes of standing waves due to the atomic
dipole interactions with the two trapping lasers at optical wavelengths
$\lambda_{x}$, $\lambda_{y}$. Thus, $V_{x,y}^{0}$ are potential
well-depths, leading to trap frequencies $\omega_{x},\,\omega_{y}$
in the $x,\, y$ directions. We have used a scale length of the reduced
oscillator lengths $d_{x,y}=\sqrt{2\hbar/m\omega_{x,y}}$ in each
direction. It is also common to use the single atom oscillator length
definition of $a_{x,y}=\sqrt{\hbar/m\omega_{x,y}}$, which we will
utilize in comparisons with experiment in later sections. 

To next order beyond the linear confinement approximation, we have
introduced $\alpha_{x},\alpha_{y}\ll1$ as the dominant anharmonic
parameters, so that:
\begin{eqnarray}
\omega_{x,y} & = & \frac{2\pi}{\lambda}\sqrt{\frac{2\left|V_{x,y}^{0}\right|}{m}}\nonumber \\
\alpha_{x,y} & = & \frac{-8\pi^{2}\hbar}{3\lambda^{2}m\omega_{x,y}}\label{eq:anharmonic_parameter}
\end{eqnarray}

For plane waves, these quartic anharmonic terms are the lowest order
possible. More generally, the potential may be neither parabolic nor
sinusoidal. Examples of this include the potential found in an optical
fibre, which can be engineered to any desired shape, and potentials
found in experiments using magnetic trapping or focused Gaussian beams.
For this reason, we expect cubic, quartic and higher order anharmonic
parameters in any real experiment. However, the quartic term given
above is due to spatial modulation on optical wavelength scales. This
is generally larger than cubic anharmonic terms like $\mathbf{r}\cdot\nabla V_{x}$
caused by focusing effects. 

For simplicity, we suppose that the the dominant anharmonic effects
are caused by anharmonic parameters $\alpha_{x,y}$, and the single-particle
Hamiltonian is: 
\begin{eqnarray}
H_{i} & = & -\frac{\hbar^{2}}{2m}\nabla_{\mathbf{r}_{i}}^{2}+\frac{1}{2}m\omega_{x}^{2}x_{i}^{2}\left(1+\alpha_{x}x_{i}^{2}/d_{x}^{2}\right)+\nonumber \\
 &  & +\frac{1}{2}m\omega_{y}^{2}y_{i}^{2}\left(1+\alpha_{y}y_{i}^{2}/d_{y}^{2}\right)\,.\,\left(i=1,2\right)\label{eq:1D 2}
\end{eqnarray}
Thus, the trapping Hamiltonian (\ref{eq:1D 1}) has the form of a
harmonic Hamiltonian $H_{h}$ plus an anharmonic terms $H_{a}^{x}$
and $H_{a}^{y}$ in the $x$and $y$ directions respectively: 
\begin{eqnarray}
H_{1}+H_{2} & = & H_{h}+H_{a}^{x}+H_{a}^{y}\,\nonumber \\
 & = & H_{h}+H_{a}\,.
\end{eqnarray}

Next, we can estimate typical parameter values in recent experiments,
as shown in Table \ref{tab:1}. We note that in the Innsbruck experiments
with relatively large observed anomalies, the dimensionless anharmonic
parameter was typically $12\%$, which is substantially larger than
in the case of the Cambridge experiment, with an anharmonicity of
$7.5\%$. These parameters are calculated for alkali metal atoms,
micron wavelength lasers and typical $10-100kHz$ trap frequencies.
Obviously, large changes in anharmonicities are easily obtained by
changing any of the relevant factors.

These anharmonic parameters lead to energy shifts in the ACIR resonances,
which we calculate below. More importantly, any type of anharmonic
potential allows a coupling between relative and COM motion, which
is otherwise prohibited due to the Kohn theorem.

\begin{table}
\centering{}%
\begin{tabular}{|c|c|c|}
\hline 
Experiment & $^{133}Cs$ & $^{40}K$\tabularnewline
\hline 
\hline 
Trap frequency $\omega$ & $2\pi\times14.5\, kHz$ & $2\pi\times80\, kHz$\tabularnewline
\hline 
Wavelength $\lambda$ & $1.064\times10^{-6}m$ & $1.064\times10^{-6}m$\tabularnewline
\hline 
Atomic mass $m$ & $2.22\times10^{-25}kg$ & $0.6635\times10^{-25}kg$\tabularnewline
\hline 
Length $d_{x,y}$ & $0.102\times10^{-6}m$ & $0.08\times10^{-6}m$\tabularnewline
\hline 
Anharmonicity $\alpha_{x,y}$ & $-0.121$ & $-0.075$\tabularnewline
\hline 
\end{tabular}\caption{Typical anharmonic parameters for CIR experiments using optical lattices,
following data from Innsbruck ($Cs$) \cite{Haller2010C} and Cambridge
(K) \cite{Frohlich2010R}. Quantitative values depend on the trap
frequency, which is varied over a range of values. \label{tab:1}}
\end{table}

\subsection{COM energies}

We can now make a preliminary estimate of atomic and molecular energies
in the two-particle sector, given the anharmonic trapping potential.
These estimates assume sufficiently tight internal binding so that
only the COM energies are changed by the anharmonicity. While this
is not accurate near threshold, it allows an estimate of the size
of anharmonic perturbation energies. It will also be used to check
the validity of subsequent results in the tight-binding limit.

\subsubsection{Atomic energy}

For a single particle, the perturbation theory solution including
the anharmonic parameter is well known. We can calculate how the free
atomic energy, 
\begin{equation}
E_{A}^{(0)}=\left(n_{x}+\frac{1}{2}\right)\hbar\omega_{x}+\left(n_{y}+\frac{1}{2}\right)\hbar\omega_{y}
\end{equation}
 is changed by the anharmonic perturbation. To first order in perturbation
theory, the modified transverse ground state energy is an elementary
perturbation theory result, such that $E_{A}=E_{A}^{(0)}+E_{A}^{(1)}\,,$where
\begin{eqnarray}
E_{A}^{(1)} & = & \left\langle \psi_{n_{x},n_{y}}\right|H_{a}\left|\psi_{n_{x},n_{y}}\right\rangle \,\nonumber \\
 & = & \frac{3}{16}\left(2n_{x}\left(n_{x}+1\right)+1\right)\hbar\omega_{x}\alpha_{x}+x\leftrightarrow y\label{eq:One-atom-anharmonic}
\end{eqnarray}
 Here, $\left|\psi_{n_{x},n_{y}}\right\rangle $ is the single-particle
eigenstate of $H_{h}$ , which is treated as the zero-order wave-function.
For a threshold resonance experiment, the incoming total energy of
two atoms initially in a transverse and longitudinal ground state
is therefore:
\begin{equation}
E_{scatt}=\hbar\omega_{x}\left(1+\frac{3}{8}\alpha_{x}\right)+\left(x\leftrightarrow y\right)\label{eq:atomic_anharmonicity}
\end{equation}

Using the numbers in Table (1), this indicates that the scale of anharmonic
energy perturbations should be around $5-10\%$ of the transverse
trap frequency, for the parameters of recent experiments.

\subsubsection{Molecular energy}

In the COM relative-coordinate frame, with $\mathbf{R}=\left(\mathbf{r}_{1}+\mathbf{r}_{2}\right)/2$
, $\mathbf{r}=\mathbf{r}_{1}-\mathbf{r}_{2}$ , the harmonic term
is:
\begin{eqnarray}
H_{h} & = & H_{h}^{CM}+H_{h}^{rel}\nonumber \\
 & = & -\frac{\hbar^{2}}{2M}\nabla_{\mathbf{R}}^{2}+\frac{1}{2}M\left(\omega_{x}^{2}X^{2}+\omega_{y}^{2}Y^{2}\right)-\nonumber \\
 &  & -\frac{\hbar^{2}}{2\mu}\nabla_{\mathbf{r}}^{2}+\frac{1}{2}\mu\left(\omega_{x}^{2}x^{2}+\omega_{y}^{2}y^{2}\right)+U\left(\mathbf{r}\right)\,,\label{eq:1D 4}
\end{eqnarray}
and the anharmonic term is:
\begin{eqnarray}
H_{a} & = & H_{a}^{x}+H_{a}^{y}\,\label{eq:1D 5}\\
 & = & \frac{\alpha_{x}m\omega_{x}^{2}}{2d_{x}^{2}}\left(2X^{4}+3X^{2}x^{2}+\frac{x^{4}}{8}\right)+x\leftrightarrow y\nonumber 
\end{eqnarray}
Here, $\mathbf{r}=\left(x,y,z\right)$ , $\mathbf{R}=\left(X,Y,Z\right)$
, and $M=2m$ , $\mu=m/2$ are the mass of the COM and the reduced
mass, respectively. 

As a point of reference for the more detailed calculations given below,
we now consider how the COM energy of molecular bound states changes
due to anharmonicity. For a broad Feshbach resonance, with strong
binding so that the internal molecular energy is not changed by the
waveguide, the internal molecular bound state energy in free space
for an attractive interaction is known to be:
\begin{equation}
E_{b}^{3D}=-\frac{\hbar^{2}}{ma_{3D}^{2}}\label{eq:3D-energy}
\end{equation}
 For a tightly bound molecule described only by the COM coordinates
$\left(X,Y\right)$, the oscillator frequencies are $\omega_{x},\,\omega_{y}$
as before. To first order in perturbation theory, the additional COM
energy $E_{CM}$ of a tightly bound molecular state is therefore:
\begin{equation}
E_{CM}=E_{CM}^{(0)}+E_{CM}^{(1)}\,,\label{eq:one-molecule-perturbation}
\end{equation}
where $E_{CM}^{(0)}=\left(N_{x}+\frac{1}{2}\right)\hbar\omega_{x}+\left(N_{y}+\frac{1}{2}\right)\hbar\omega_{y}$,
and:
\begin{eqnarray}
E_{CM}^{(1)} & = & \left\langle \Psi_{N_{x}N_{y}}\right|H_{a}\left|\Psi_{N_{x}N_{y}}\right\rangle \,\nonumber \\
 & = & \frac{3\hbar\omega_{x}\alpha_{x}}{32}\left[1+2N_{x}\left(N_{x}+1\right)\right]+x\leftrightarrow y\label{eq:one-molecule-anharmonic}
\end{eqnarray}
 Here, $\left|\psi_{N_{x},N_{y}}\right\rangle $ is the single-molecule
eigenstate of $H_{h}$, again treated as the zero-order wave-function,
and $E_{CM}^{(0)}$, $E_{CM}^{(1)}$ are the harmonic and anharmonic
contributions respectively to the COM molecular energy . The reduced
anharmonic correction compared to the free atomic case is due to the
reduced spatial width of the wave-function, caused by the increased
molecular mass compared to an atom.

\subsection{Tightly bound resonance threshold}

This allows us to make a relatively simple calculation. A threshold
condition in the tight-binding limit is obtained from equating the
total ground state transverse energy of two atoms with the total molecular
energy in an COM excited transverse state. It will be convenient for
later calculations to define a dimensionless bound state energy relative
to unbound atoms in a transverse waveguide as: 
\begin{equation}
\epsilon\equiv\frac{E_{b}}{\hbar\omega_{y}}-\frac{1}{2}\left(1+\eta\right)\,,\label{eq:dimensionless_binding_energy}
\end{equation}
where $\eta=\omega_{x}/\omega_{y}$ is the anisotropy of the two transverse
binding frequencies, and $E_{b}=E_{b}^{3D}$ in the strong binding
limit of interest here. The correction term of $-\frac{1}{2}\left(1+\eta\right)$
is required to take account of the difference in the transverse confinement
energies between the atoms and the molecular state, which does not
occur in free space. 

After including the anharmonic corrections and excitation energies
from Eq (\ref{eq:One-atom-anharmonic}) and Eq (\ref{eq:one-molecule-anharmonic}),
one obtains a resonance condition for \emph{deeply} bound molecular
ACIR in an intuitive form as:

\begin{equation}
E_{scatt}=E_{b}+E_{CM}
\end{equation}
On transforming this to dimensionless form, we obtain:
\begin{eqnarray}
\epsilon+N_{y}+\alpha_{y}\left[\frac{3}{16}N_{y}\left(N_{y}+1\right)-\frac{9}{32}\right]+\nonumber \\
+\eta\left\{ N_{x}+\alpha_{x}\left[\frac{3}{16}N_{x}\left(N_{x}+1\right)-\frac{9}{32}\right]\right\}  & = & 0\,.\label{eq:deeply bound}
\end{eqnarray}

Clearly there are multiple resonances as $N_{x}$ and $N_{y}$ is
varied, thus altering the COM quantum numbers of the excited transverse
molecular states. The position of these resonances is largely determined
by the quantum numbers, together with anharmonic shifts. In the next
section, we show that wave function symmetries mean that the even
order resonances are directly coupled by the strong quartic anharmonicities
$\alpha_{x},\alpha_{y}$. Odd order resonances are coupled through
the relatively weaker cubic anharmonic terms due to the $\mathbf{r}\cdot\nabla V_{x}$
terms, which are physically caused by the slow variations in the Gaussian
envelope function of the trapping lasers in these experiments. 

\begin{figure}
\includegraphics[width=1\columnwidth]{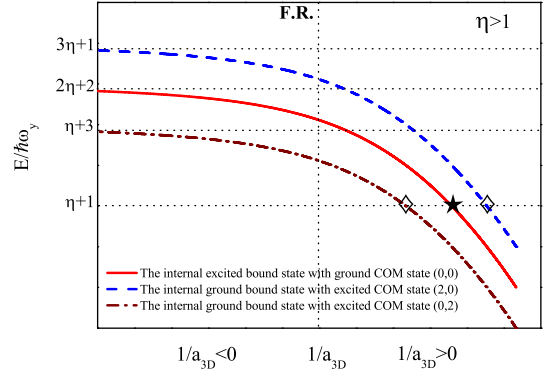}\caption{(Color online) Diagram of the confinement induced energy levels in
an anisotropic trap or waveguide, with anisotropy $\eta$. The solid
line is an internally excited two-particle ground state. The dashed
lines correspond to a COM excitations of the two-particle ground state.
Confinement induced resonance thresholds occur when these excited
molecular states become resonant with two incoming free particles
with zero momentum. This is indicated by a star in the internally
excited CIR case, which is the only level coupled in the case of a
harmonic trap. For anharmonic traps, these additional levels become
coupled to input atomic states with zero transverse excitation, and
acquire additional anharmonic energy shifts. \label{fig:(Color-online)-Energy-Levels-1}}

\label{Flo:Innsbruck-1}
\end{figure}

The consequences are seen in Fig.\ref{fig:(Color-online)-Energy-Levels-1},
which shows the first two strongly coupled even-order COM resonances,
as compared with the internally excited resonance position. The traditional
harmonic confinement induced resonance (HCIR) state in a one dimensional
harmonic waveguide is identified with the solid line in Fig.\ref{fig:(Color-online)-Energy-Levels-1}.
This curve is simply the first \emph{internally} excited state of
the two-body ground state \cite{Olshanii1998A,Peng2010C}. One expects
a CIR to occur whenever the solid curve crosses the lowest horizontal
line, thus permitting a resonance to occur with incoming atoms near
zero energy. 

However, we see that there are also two further possibilities, spaced
both above and below the internally excited resonance. These are the
lowest lying even order ACIR resonances, which we expect to be the
dominant excitations in the case of anharmonic waveguides.

While this calculation is approximately correct, and gives an excellent
intuitive picture, we will carry out a quantitative calculation in
the following sections with the inclusion of the internal degrees
of freedom as well.

\section{Anharmonic CIR in a quasi-1D waveguide\label{sec:Nonlinear-CIR-1D}}

While Eq (\ref{eq:deeply bound}) is valid for a relatively deeply
bound molecule, it neglects both anharmonic and waveguide corrections
to the \emph{internal} molecular energy $E_{b}$. The question of
which transverse molecular quantum numbers $N_{x},$$N_{y}$ are accessible
in terms of selection rules also needs to be addressed more carefully.
In this section, we therefore treat the general case of a quasi-1D
confining waveguide with an asymmetric confining potential. This case
can be continuously changed to the limit of the 2D trap, which is
treated in more detail in the following section. Due to the small
anharmonic parameter $\alpha$, the anharmonic term $H_{a}$ will
be treated as a perturbation to $H_{h}$ .

\subsection{Harmonic CIR }

Firstly, we consider the anisotropic waveguide \emph{without} the
anharmonic perturbation $H_{a}$. While this has been treated previously\cite{Peng2010C},
we revisit it here as a first step to solving the anharmonic case.
The origin of the confinement induced resonance is due to two scattering
atoms forming a virtual molecule via their S-wave interaction. The
energy of the resulting two atom quasi-bound state in a waveguide
is written as: 
\begin{eqnarray}
E_{M}^{(0)} & = & E_{CM}+E_{b}\nonumber \\
 & = & \left(N_{x}+\frac{1}{2}\right)\hbar\omega_{x}+\left(N_{y}+\frac{1}{2}\right)\hbar\omega_{y}+E_{b}\label{eq:8}
\end{eqnarray}
 Here $E_{CM}$ is the energy of the COM excitation, $\left(N_{x},N_{y}\right)$
are the quantum numbers of the COM, and $E_{b}$ is the binding energy
of the two atoms. This becomes resonant with two incoming atoms near
zero momentum when $E_{M}^{(0)}=E_{scatt}$, where $E_{scatt}$ is
the incoming free-particle energy. This is trivially given in the
zero-momentum, harmonic case by:
\begin{eqnarray}
E_{scatt} & = & 2E_{A}^{(0)}\nonumber \\
 & = & \hbar\omega_{x}+\hbar\omega_{y}\,.
\end{eqnarray}

By solving the eigenproblem of the relative Hamiltonian of the two
atoms $H_{h}^{rel}$ , we can obtain the relation between the binding
energy $E_{b}$ and the 3D s-wave scattering length $a_{3D}$. This
is known from previous work \cite{Peng2010C}, by solving for the
dimensionless energy $\epsilon$ of the molecular ground state, where
$\epsilon\equiv E_{b}/\hbar\omega_{y}-\left(\eta+1\right)/2$, as
given in Eq (\ref{eq:dimensionless_binding_energy}). We note that,
just as with Eq (\ref{eq:dimensionless_binding_energy}) in the previous
section, the dimensionless energy $\epsilon$ is defined so that it
includes the change in transverse confinement energies. With this
definition, $E_{b}$ reduces to the free-space binding energy in the
limit of weak confinement or strong binding. 

The dimensionless ground state molecular energy $\epsilon$ is given
by an implicit equation \cite{Peng2010C}:
\begin{equation}
\frac{d_{y}}{a_{3D}}=-\frac{1}{\sqrt{\pi}}\mathcal{F}_{1}\left(\epsilon,0\right)\,,\label{eq:dimensionless_energy}
\end{equation}
 where the RHS is defined by the definite integral 
\begin{equation}
\mathcal{F}_{1}\left(\epsilon,0\right)=\int_{0}^{\infty}dt\left[\frac{\sqrt{\eta}\exp\left(\epsilon t/2\right)}{\sqrt{t\left(1-e^{-\eta t}\right)\left(1-e^{-t}\right)}}-\frac{1}{t^{3/2}}\right]\,.\label{eq:1D 9}
\end{equation}

In the strong binding limit, the limiting behavior of this integral
is:
\begin{equation}
\lim_{\epsilon\rightarrow-\infty}-\frac{\mathcal{F}_{1}\left(\epsilon,0\right)}{\sqrt{\pi}}=\sqrt{\frac{2\left|E_{b}\right|}{\hbar\omega_{y}}}.
\end{equation}
This is leads to the free-space three-dimensional binding energy result,
Eq (\ref{eq:3D-energy}), as one expects in this limit. Combining
Eq (\ref{eq:dimensionless_binding_energy}) and Eq (\ref{eq:8}),
the threshold condition for a \emph{harmonic} trap can be summarized
compactly in one equation as 
\begin{equation}
\epsilon+N_{x}\eta+N_{y}=0\,,\label{eq:Harmonic threshold}
\end{equation}
 so that:
\begin{equation}
\frac{d_{y}}{a_{3D}}=-\frac{1}{\sqrt{\pi}}\mathcal{F}_{1}\left(-\left[N_{x}\eta+N_{y}\right],0\right)\,.\label{eq:dimensionless_energy-1}
\end{equation}

Further resonances are anticipated if we consider relative atomic
motion. Such an internally excited molecular state is described by
a completely different integral equation. For the first excited state
of the internal motion, one obtains:

\begin{equation}
\frac{d_{y}}{a_{3D}}=-\frac{1}{\sqrt{\pi}}\mathcal{F}_{e}\left(\epsilon_{e},0\right)\,,\label{eq:dimensionless_energy-2}
\end{equation}
 where the RHS is now defined by the definite integral 
\begin{eqnarray}
\mathcal{F}_{e}\left(\epsilon_{e},0\right) & = & \int_{0}^{\infty}dt\biggl[e^{\epsilon_{e}t/2}\sqrt{\frac{\eta}{t}}\biggl(\frac{1}{\sqrt{\left(1-e^{-\eta t}\right)\left(1-e^{-t}\right)}}\nonumber \\
 &  & -1\biggr)-\frac{1}{t^{3/2}}\biggr]\,.
\end{eqnarray}

The CIR threshold condition including both internal and COM excitations
is therefore 
\begin{equation}
\epsilon_{e}+N_{x}\eta+N_{y}=0\,,\label{eq:internal_threshold_harmonic}
\end{equation}
so that:
\begin{equation}
\frac{d_{y}}{a_{3D}}=-\frac{1}{\sqrt{\pi}}\mathcal{F}_{e}\left(-\left[N_{x}\eta+N_{y}\right],0\right)\,.\label{eq:dimensionless_energy-1-1}
\end{equation}

The above equations \ref{eq:Harmonic threshold} and \ref{eq:internal_threshold_harmonic}
give threshold conditions in the limit of small anharmonicity, for
coupling to either the internal ground or excited state respectively,
with center-of-mass quantum numbers included in the final resonant
state. 

As such, they give an elegant picture of the possible resonances,
including both internal and COM excitations. However, if the anharmonic
term $H_{a}$ is not included, the two incoming atoms in a transverse
ground state cannot couple to the transverse excited molecular states
during the collision. Hence, for harmonic confinement there is only
one observable CIR resonance no matter how anisotropic the transverse
confinement \cite{Peng2010C}. This is described by the last equation
above, Eq (\ref{eq:dimensionless_energy-1-1}), on setting $N_{x}=N_{y}=0$. 

In reality, anharmonic terms do occur. These lead both to couplings
that allow COM excitations, and to energy shifts which alter the resonance
locations. In the following analysis, we will assume that there are
only COM excitations, and we will apply perturbation theory to the
bound state COM energies predicted by Eq (\ref{eq:Harmonic threshold}).

\subsection{Anisotropic, anharmonic CIR }

If the anharmonic perturbation $H_{a}$ is now introduced, the COM
motion will be mixed with the relative motion by the anharmonicity
of the confining trap. Then the transversely excited COM molecular
states can couple to the scattering state of the two incoming atoms
in the transverse ground state. However, both the atomic and molecular
states now have energy levels shifted by anharmonic corrections. This
means that the fundamental resonance equation is modified from Eq
(\ref{eq:Harmonic threshold}) for the harmonic trap case. It is now:
\begin{equation}
\epsilon+N_{x}\eta+N_{y}+\epsilon_{a}\left(N_{x},N_{y}\right)=0\,,\label{eq:Harmonic threshold-1}
\end{equation}
where $\epsilon_{a}\left(N_{x},N_{y}\right)$ is the anharmonic correction
to the relative energy levels for a COM excitation with quantum numbers
$\left(N_{x},N_{y}\right)$. This has been treated already in the
deeply bound limit in Eq (\ref{eq:deeply bound}). We now treat this
in the general case. 

From Eq (\ref{eq:One-atom-anharmonic}), one must include the input
anharmonicity in the atomic levels, which for the case of two atoms
in an initial atomic transverse ground state is given by Eq (\ref{eq:atomic_anharmonicity}).
Next, we consider the effects of anharmonicity on the bound or molecular
energies. 

With a quartic anharmonicity which is symmetric around the origin,
there are constraints on the types of coupling that can occur to the
COM motion. In particular, with a symmetric input state having $N_{x}=N_{y}=0$,
one must have a symmetric resonance state. This implies that only
even COM quantum numbers are strongly coupled. There is also a weak
coupling to odd COM quantum numbers, caused by cubic anharmonic parameters,
which we treat in the next section. The lowest of the strong nonlinear
resonances occur when $\left(N_{x},N_{y}\right)=\left(2,0\right)$
or $\left(0,2\right)$. Consequently the resonance splits if the transverse
confinement is anisotropic. 

We give a detailed calculation of the effects of anharmonic confinement
on the bound-state energies of these resonances in the Appendix. Using
these results, we arrive at the resonance condition for the $N_{x}=2$
state ,
\begin{eqnarray}
\epsilon+2\eta+\alpha_{x}\eta\left(\frac{27}{32}-\frac{5\eta}{32\epsilon}+\frac{3\eta^{2}}{320\epsilon^{2}}\right)\nonumber \\
+\alpha_{y}\left(\frac{-9}{32}-\frac{1}{32\epsilon}+\frac{3}{320\epsilon^{2}}\right) & = & 0\label{eq:Nx_general}
\end{eqnarray}
 In like manner, the resonance condition for the $N_{y}=2$ state
is,
\begin{eqnarray}
\epsilon+2+\alpha_{x}\eta\left(\frac{-9}{32}-\frac{\eta}{32\epsilon}+\frac{3\eta^{2}}{320\epsilon^{2}}\right)\nonumber \\
+\alpha_{y}\left(\frac{27}{32}-\frac{5}{32\epsilon}+\frac{3}{320\epsilon^{2}}\right) & = & 0\label{eq:Ny_general}
\end{eqnarray}

These results can be compared with Eq (\ref{eq:deeply bound}), which
were obtained in the previous section, dealing with the case of a
deeply bound molecular state. On dropping terms scaling with $1/\epsilon$,
the two conditions agree in the limit of strong molecular binding,
with $\epsilon\rightarrow-\infty$. 

For experiments using optical lattices with equal wavelengths in each
direction, one finds that:
\begin{equation}
\alpha_{x}\eta=\alpha_{y}\equiv\alpha
\end{equation}
 Then the resonance conditions can be reduced to a simpler form. For
$\left(N_{x},N_{y}\right)=\left(2,0\right)$, one obtains
\begin{equation}
\epsilon+2\eta+\alpha\left(\frac{9}{16}-\frac{5\eta+1}{32\epsilon}+\frac{3\left(\eta^{2}+1\right)}{320\epsilon^{2}}\right)=0\,,\label{eq:N_x_equation}
\end{equation}
 and for $\left(N_{x},N_{y}\right)=\left(0,2\right)$ :
\begin{equation}
\epsilon+2+\alpha\left(\frac{9}{16}-\frac{\eta+5}{32\epsilon}+\frac{3\left(\eta^{2}+1\right)}{320\epsilon^{2}}\right)=0\,.\label{eq:N_y_excitations}
\end{equation}

\begin{figure}
\includegraphics[width=1\columnwidth]{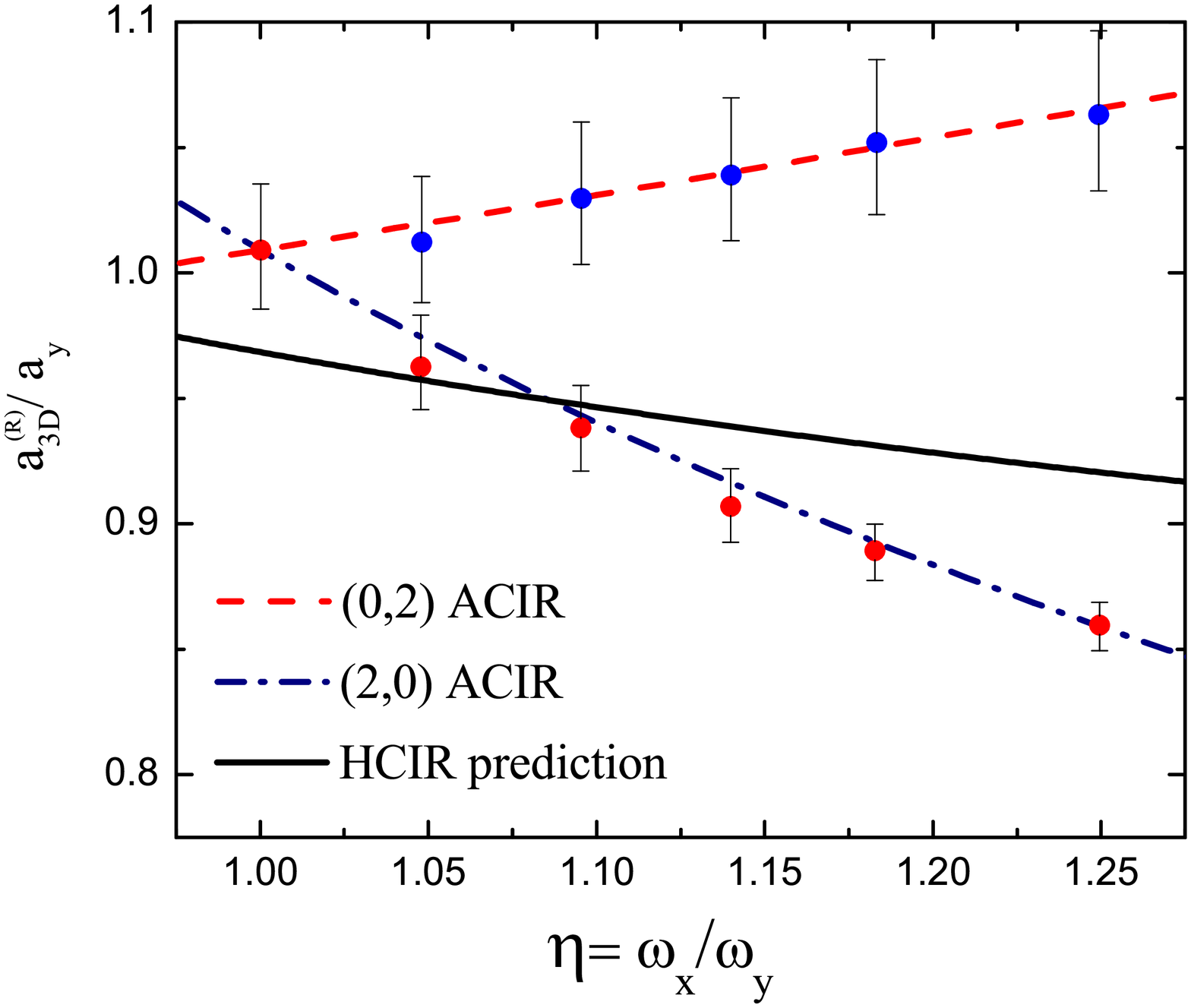}

\caption{(Color online) The predictions of ACIR resonances using  first-order
anharmonic perturbation theory comparing with the experimental data
\cite{Haller2010C}. The black solid curve is the result of a generalized
Olshanii's HCIR model \cite{Peng2010C}. Here the anisotropy $\eta$
is changed by increasing the frequency $\omega_{x}$ with a fixed
$\omega_{y}=2\pi\times13.2(2)$kHz (or $\omega_{1}$ in\cite{Haller2010C}),
which resulted in an anharmonic parameter of $\alpha=-0.133$ . \label{fig:The-predictions-unshifted}}
\end{figure}

Solving these equations requires the use of a nonlinear equation solving
numerical algorithm, like the Newton-Raphson method which we use in
Fig.\ref{fig:The-predictions-unshifted} to obtain theoretical results
for comparison with experiment.

Now we are ready to understand the dominant effects observed in the
recent Innsbruck experiment\cite{Haller2010C}. The atoms are prepared
in a 1D tube geometry using two orthogonal pairs of counter-propagating
laser fields to create a trapping environment. In the case of transverse
isotropic confinement, due to the small collision energy and large
energy interval between the transverse energy levels, only the first
excited molecular states $\left(N_{x}=2,N_{y}=0\right)$ and $\left(N_{x}=0,N_{y}=2\right)$
of the COM motion which are degenerate can couple to the scattering
state of two incoming atoms. In this isotropic situation, the results
of ACIR are similar to those of HCIR and also agree well with the
experiment, as long as the anharmonicity of the confinement is not
too large. Due to the broad resonances and fitting techniques used,
the isotropic experimental data has too little precision to accurately
distinguish between ACIR and CIR resonances.

By increasing the transverse anisotropy $\eta=\omega_{x}/\omega_{y}$
, the molecular states $\left(N_{x}=2,N_{y}=0\right)$ and $\left(N_{x}=0,N_{y}=2\right)$
of the COM motion are no longer degenerate. A splitting of the ACIR
is expected, which has been observed in the experiment. By contrast,
no splitting of the linear HCIR resonance is predicted. Thus, the
double resonance observed experimentally is a clear signature of the
ACIR resonance. It is also a sign that if there is any internally
excited HCIR effect in this experiment, it is relatively suppressed
compared to the ACIR effect which involves COM excitations.

The theoretical predictions of the CIRs compared with the experimental
results \cite{Haller2010C} are presented in Fig.\ref{fig:The-predictions-unshifted}.
As we can see, the predictions of the COM-relative coupling theory
are consistent with the experimental data. 

In these experiments, the resonances were identified as occurring
at the point of maximum molecular loss. This leads to an offset between
the apparent and true resonance due to finite resonance widths. To
compensate for this in the original experimental publication, the
data was fitted to the isotropic HCIR prediction by adding a small
constant offset to make it equal to the theoretical results at $\eta=1$.
We follow a similar fitting procedure here, for consistency, but we
fit the offset to the isotropic ACIR prediction instead. Note that
the anharmonic parameter is different to that used in Table \ref{tab:1},
because Table \ref{tab:1} gives values representative of a range
of experiments, while in each of the figures we use the data from
the relevant experiment.

At the point $\eta=1$, there is transversely \emph{isotropic} confinement,
and the ACIR prediction is in accord with the Olshanii CIR, apart
from anharmonic corrections. However, as the transverse confinement
becomes more and more anisotropic, the internally excited HCIR resonance
persists as a single resonance except for a small frequency shift
\cite{Peng2010C}. Both ACIR theory and experiment show a clear splitting
of the original resonance with increasing trap asymmetry. There is
excellent quantitative agreement in the amount of splitting. In these
experiments, there is strong evidence for the COM excited ACIR resonances.

\subsection{Multiple resonance ACIR at large anisotropy}

As the transverse anisotropy $\eta$ becomes even larger, the energy
spacing in the $y$ direction decreases, and more transversely excited
molecular states can readily couple to the initial scattering state.
Consequently, more additional COM resonances can occur. By contrast,
there is still only one internal HCIR resonance predicted no matter
how large the transverse anisotropy. However, several multiple resonances
are observed experimentally at large anisotropy. In order to understand
these, we now consider the other transverse states.

As in the previous section, given any coupling parameter ($a_{3D}$),
the unperturbed binding energy $\epsilon$ is determined by Eq (\ref{eq:dimensionless_energy}),
which as an implicit equation involving the integral $\mathcal{F}_{1}\left(\epsilon,0\right)$.
We then use perturbation theory to calculate the dimensionless anharmonic
energy shift $\epsilon_{a}$. Hence, we obtain the ACIR resonance
position of $a_{3D}$ for arbitrary odd and even COM quantum numbers
from the solutions to the overall resonance equation, Eq (\ref{eq:Harmonic threshold-1}).

For brevity, we will refer to an arbitrary molecular COM resonance
as simply $\left(N_{x},N_{y}\right)$. The general form of the resulting
anharmonic shifts is derived in the Appendix, for arbitrary quartic
anharmonic parameters. We include both odd and even order resonances,
because, as remarked earlier, there are both cubic and quartic anharmonic
couplings, which leads to the possibility of both even and odd ACIR
resonances. However, for simplicity we do not include the relatively
small cubic energy shifts.

For comparison to current experiments, we are interested in the case
of equal optical trapping wavelengths, which means that $\alpha_{x}\eta=\alpha_{y}\equiv\alpha$.
From the Appendix, the general form of the resulting anharmonic shifts
for resonances corresponding to the $\left(N_{x},N_{y}\right)$ COM
state is as follows,
\begin{multline}
\epsilon_{a}\left(N_{x},N_{y}\right)/\alpha=\frac{3(N_{y}^{2}+N_{y}+N_{x}^{2}+N_{x})-9}{16}-\\
-\frac{2N_{y}+1+\eta(2N_{x}+1)}{32\epsilon}+\frac{3\left(\eta^{2}+1\right)}{320\epsilon^{2}}
\end{multline}

This allows the positions of ACIR denoted by $a_{3D}$ to be calculated.
In the experimental reports of multiple resonances at large anisotropy
(Fig.4 \cite{Haller2010C}), the loss rates are plotted as a function
of magnetic field, rather than $a_{3D}$. We therefore make use of
the relation between the 3D scattering length $a_{3D}$ and $B$-field
for the relevant $^{137}Cs$ Feshbach resonance \cite{Lange2009D},
which is,
\begin{equation}
\frac{a_{3D}}{a_{bg}}=\frac{B-18.1}{B+11.1}\cdot\frac{B-47.944}{B-47.78}\cdot\frac{B-53.457}{B-53.449}\label{eq:3}
\end{equation}
 and,
\begin{eqnarray}
\frac{a_{3D}}{a_{bg}} & = & \frac{1}{1875a_{0}}\sqrt{\frac{\hbar\eta}{m\omega_{x}}}\cdot\frac{a_{3D}}{a_{y}}\nonumber \\
 & = & 0.681\sqrt{\eta}\cdot\frac{a_{3D}}{a_{y}}
\end{eqnarray}

Hence, the predicted magnetic field at resonance can be calculated.
In order to compare our theory to these experiments, the regime of
anisotropy $\eta$ considered is $\left[1.4,2.3\right]$ . We also
note that in these experiments, the trapping frequency $\omega_{y}$
is varied, so the effective anharmonicity parameter $\alpha$ therefore
changes at different anisotropy.

\begin{figure}
\includegraphics[width=1\columnwidth]{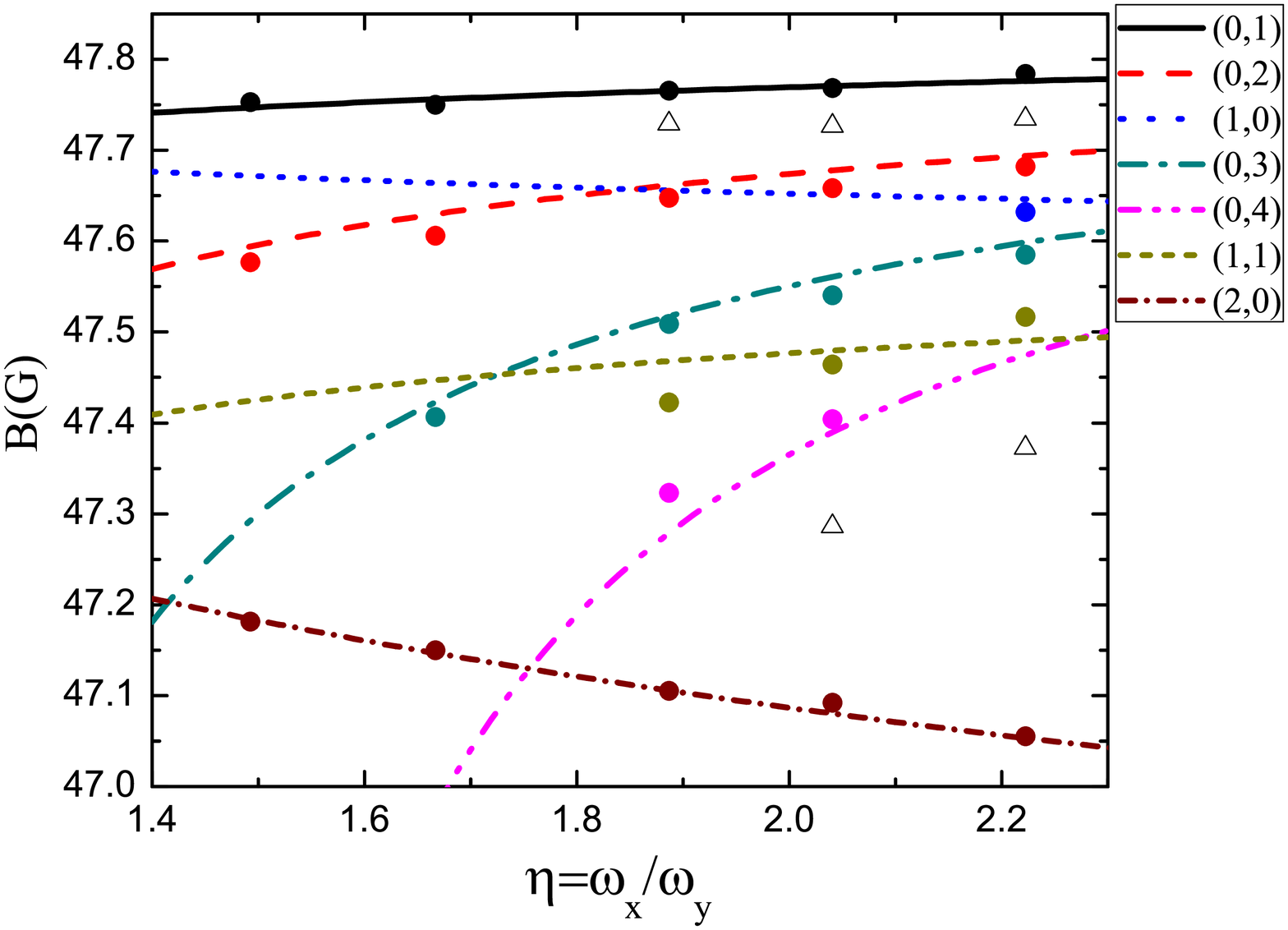}

\caption{(Color online) The resonant magnetic field $B$ vs anisotropy $\eta$
at ACIR resonances (solid and dotted lines identified alongside the
figure) from first-order anharmonic perturbation theory comparing
with the Innsbruck experiment data (round circles). Data is extracted
from multiple resonance scans \cite{Haller2010C} at absorption edges.
Here the anisotropy $\eta$ is changed by decreasing the frequency
$\omega_{y}$ with a fixed $\omega_{x}=2\pi\times16.6(2)$kHz which
results in an $\eta$-dependent anharmonic parameter $\alpha=-0.106\eta$
. \label{fig:Multiple-resonance}}

\label{fig4}
\end{figure}

Then main results are summarized in Fig.\ref{fig4}, which compares
theory to experiment. The experimental resonance points are obtained
from the raw data as the start of the resonance edges, which is appropriate
for this type of resonance experiment. No fitting parameters or shifts
are employed. Since error bars were not given in the experimental
plot, we are unable to estimate these.

Most observed resonances can be easily identified, which are coded
in the same color as the corresponding theoretical predictions. There
are $25$ identified resonances, all of which are in excellent agreement
with theoretical calculations. However, there are $5$ smaller resonances
not identified, which are indicated by the empty triangles. 

Apart from experimental issues, possible explanations for these unidentified
peaks are:
\begin{itemize}
\item \emph{Higher-order many-body ACIR resonances.} Combination 4-body
cluster resonances could occur at intermediate points between the
identified 2-body resonances. For example, the three unidentified
resonances between $(0,1)$ and $(0,2)$ could be caused by the simultaneous
excitation of $(0,1)$ and $(0,2)$ in a 4-body collision. Similarly,
the three resonances we have identified as $(1,1)$ resonances could
also be caused by 4-body excitation of $(0,3)$ and $(0,4)$ ACIR
resonances.
\item \emph{Anharmonically shifted internal HCIR resonances.} We have not
calculated these, as the anharmonic shifts of these resonances are
outside the scope of this paper. However, this mechanism provides
a possible explanation for the two unidentified resonances between
$(0,4)$ and $(2,0)$. 
\end{itemize}

\section{Anharmonic CIR in a quasi-2D system\label{sec:Nonlinear-CIR-2D}}

For a quasi-2D system, atoms are tightly confined in the axial direction,
and can freely move in two transverse directions. In the following
section, we will derive the anharmonic CIR resonance properties in
this case as well. As we will show, these results can be also obtained
from results in the previous section by taking one of the confinement
frequencies to zero. However, the direct calculation given here is
an important check on the consistency of our approach. In the case
of an optical lattice, the trapping potential is of form:
\begin{eqnarray}
U^{ext}\left(\mathbf{r}\right) & = & V_{z}\cos^{2}\left(\frac{2\pi z}{\lambda}\right)\\
 & \approx & \frac{1}{2}m\omega^{2}z^{2}\left(1+\frac{\alpha z^{2}}{d^{2}}\right)\nonumber 
\end{eqnarray}

Here, $V_{z}$ is the potential well-depth with one standing-wave
trapping laser at optical wavelength $\lambda$ . This leads to the
trap frequency $\omega$ in the $z$ direction. We note that cubic
anharmonic terms are also possible due to focusing effects in this
case also.

As before, we use a scale length of the reduced oscillator length
$d=\sqrt{2\hbar/m\omega}$ , and define a single atom oscillator length
$a_{\perp}=\sqrt{\hbar/m\omega}$, and a small anharmonic parameter
$\alpha$, so that for an optical lattice,
\begin{equation}
\alpha=\frac{-8\pi^{2}\hbar}{3\lambda^{2}m\omega}\,.
\end{equation}

The Hamiltonian of two atoms in a quasi-2D system is then,
\begin{equation}
H=H_{1}+H_{2}+U\left(\mathbf{r}_{1}-\mathbf{r}_{2}\right)\,,\label{eq:1}
\end{equation}
 where,
\begin{equation}
H_{i}=-\frac{\hbar^{2}}{2m}\nabla_{\mathbf{r}_{i}}^{2}+\frac{1}{2}m\omega^{2}z_{i}^{2}\left(1+\alpha\frac{z_{i}^{2}}{d^{2}}\right)\qquad\left(i=1,2\right)\,.\label{eq:2}
\end{equation}
As previously, $U\left(\mathbf{r}_{1}-\mathbf{r}_{2}\right)$ is the
interatomic interaction, and we set 
\begin{eqnarray}
\mathbf{R} & = & \frac{1}{2}\left(\mathbf{r}_{1}+\mathbf{r}_{2}\right)\nonumber \\
\mathbf{r} & = & \mathbf{r}_{1}-\mathbf{r}_{2}\,.
\end{eqnarray}
 Then Eq.(\ref{eq:1}) becomes,
\begin{equation}
H=H_{h}+H_{a}\label{eq:5}
\end{equation}
 Where,
\begin{eqnarray}
H_{h} & = & \left(-\frac{\hbar^{2}}{2M}\nabla_{\mathbf{R}}^{2}+\frac{1}{2}M\omega^{2}Z^{2}\right)+\nonumber \\
 &  & +\left[-\frac{\hbar^{2}}{2\mu}\nabla_{\mathbf{r}}^{2}+\frac{1}{2}\mu\omega^{2}z^{2}+V\left(\mathbf{r}\right)\right]\nonumber \\
 & = & H_{h}^{CM}+H_{h}^{rel}\label{eq:6}
\end{eqnarray}
\begin{eqnarray}
H_{a} & = & \alpha\cdot\frac{m\omega^{2}}{2d^{2}}\left(2Z^{4}+3Z^{2}z^{2}+\frac{1}{8}z^{4}\right)\label{eq:7}
\end{eqnarray}
 Here $\mathbf{r}=\left(x,y,z\right)$ , $\mathbf{R}=\left(X,Y,Z\right)$
, and $M=2m$ , $\mu=m/2$ .

Due to the small anharmonic parameter $\alpha$ , the anharmonic term
$H_{a}$ can be treated as a perturbation to $H_{h}$ . Firstly, consider
the case without the small perturbation $H_{a}$ . An anharmonic confinement-induced
resonance is expected when the energy of the virtual molecule is degenerate
with the energy of two incoming atoms from the axial ground state,
\begin{equation}
E_{M}^{(0)}=E_{CM}+E_{b}=E_{scatt}
\end{equation}
Where $E_{CM}=\left(N+\frac{1}{2}\right)\hbar\omega$, $E_{scatt}=\hbar\omega$,
and the binding energy $E_{b}$ is determined by,
\begin{equation}
\frac{d}{a_{3D}}=-\frac{1}{\sqrt{\pi}}\mathcal{F}_{2}\left(\epsilon,0\right)\label{eq:9}
\end{equation}
and the integral expression required here is given by,
\begin{equation}
\mathcal{F}_{2}\left(\epsilon,0\right)=\int_{0}^{\infty}dt\left[\frac{\exp\left(\frac{1}{2}\epsilon t\right)}{t\sqrt{1-e^{-t}}}-\frac{1}{t^{3/2}}\right]\label{eq:10}
\end{equation}
We define $E_{b}=\left(\epsilon+1/2\right)\hbar\omega$ , $d=\sqrt{\hbar/\mu\omega}$
. The lowest resonance occurs as $N=2$; however, this is not coupled
to the incoming states unless there is an anharmonic term in the potential. 

If the anharmonic term $H_{a}$ is included, we need to consider how
the energy of the input atomic states and virtual molecule $E_{M}$
is affected by this perturbation.

\subsection{The anharmonic energy $E_{M}^{(1)}$ }

Including the first order modification $E_{M}^{(1)}$ , we arrive
at the following integral equation for $\epsilon$,
\begin{equation}
\left(N+\frac{1}{2}\right)\hbar\omega+\left(\epsilon+\frac{1}{2}\right)\hbar\omega+E_{M}^{(1)}\left(\epsilon\right)=\hbar\omega+\frac{3}{8}\alpha\hbar\omega\label{eq:21-1}
\end{equation}
By solving this equation, the binding energy $\epsilon^{(R)}$ at
the resonance is obtained. Then substituting $\epsilon^{(R)}$ into
the Eq.(\ref{eq:9}), we will obtain the 3D scattering length $a_{3D}^{(R)}$
at the resonance. Following the general procedure outlined in the
Appendix, we find that a resonance occurs at:

\begin{equation}
N+\epsilon+\frac{\alpha}{32}\left[6N\left(N+1\right)-9-\frac{\left(2N+1\right)}{\epsilon}+\frac{3}{10\epsilon^{2}}\right]=0
\end{equation}
Hence, we obtain the equation that $\epsilon$ should satisfy for
states with $N=2$ ,
\begin{equation}
\epsilon+2+\alpha\left(\frac{27}{32}-\frac{5}{32\epsilon}+\frac{3}{320\epsilon^{2}}\right)=0\label{eq:31}
\end{equation}

This result is identical to Eq (\ref{eq:Ny_general}) in the limit
of $\eta\rightarrow0$, as one might expect, since the quasi-1D trap
becomes two-dimensional in this limit. However, it is instructive
that one cannot regain this limit directly from Eq (\ref{eq:N_y_excitations}),
which holds for optical lattices with equal wavelengths. The reason
is very simple: in an optical lattice at low transverse confinement
frequency, the transverse wave-function becomes more and more deconfined.
This increases the relative anharmonicity, as given in Eq (\ref{eq:anharmonic_parameter}),
so that $\alpha\sim1/\omega$. Therefore, our anharmonic perturbation
theory would break down for a weakly confined 1D system described
by Eq (\ref{eq:N_y_excitations}). Optical lattices that are only
weakly confining require a full Bloch wave-function theory, typically
requiring detailed numerical diagonalization \cite{Buchler}. 

\begin{figure}
\includegraphics[width=0.5\textwidth]{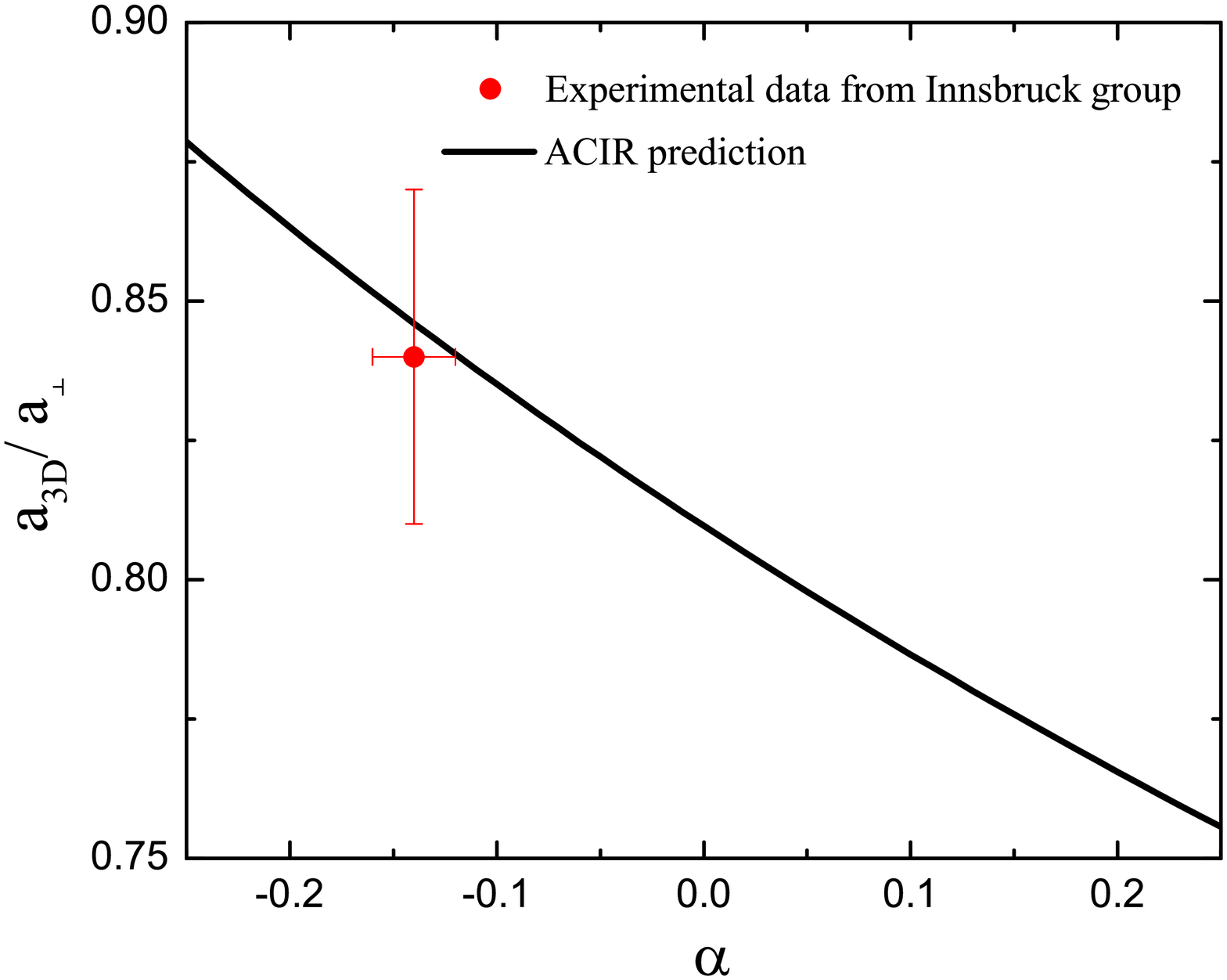}

\caption{(Color online) Predicted resonance positions of 2D CIRs, as a function
of the dimensionless anharmonic parameter $\alpha$. The experimental
data point is an average from experiments \cite{Haller2010C}, plotted
at the average anharmonicity value.\label{fig:2DResonance_vs_alpha}}
\end{figure}

From Eq.(\ref{eq:31}), we can calculate the relative binding energy
$\epsilon$ for the states of $N=2$, in the two-dimensional limit.
Then, substituting into Eq (\ref{eq:9}), the corresponding resonance
scattering length $a_{3D}^{(R)}$ is obtained. Here, in order to calculate
$\epsilon$ numerically, we use the Newton-Raphson method as before,
together with a numerical calculation of $\mathcal{F}_{2}\left(\epsilon,0\right)$.
The position of the ACIR denoted by the 3D scattering length as a
function of the anharmonic parameter $\alpha d^{2}$ is presented
in Fig.\ref{fig:2DResonance_vs_alpha}, and we can see that the 2D
ACIR is predicted at the regime of $a_{3D}>0$ . This is in contrast
to the internal HCIR resonance, which occurs for the attractive regime
with $a_{3D}<0$. 

Now we can compare these predictions with 2D resonances observed in
the recent Innsbruck experiment\cite{Haller2010C}. When one of the
lattice lasers is turned off, the system approaches a 2D geometry.
All resonances disappear except one at $a_{3D}>0$ . Given the average
anharmonicity in this experiment of $\alpha=-0.121$, we expect as
shown in Fig.\ref{fig:2DResonance_vs_alpha} to find a resonance at
\begin{equation}
a_{3D}=0.85a_{\perp}
\end{equation}

The observed resonances occur at a constant ratio between $a_{3D}$
and $a_{\perp}$, such that 
\begin{equation}
\frac{a_{3D}}{a_{\perp}}a_{3D}=0.84(3)
\end{equation}
This is consistent with the prediction of 2D ACIR. The anharmonicity
$\alpha$ in the experiments was varied through a small range as the
trapping frequency varied, and we have calculated the ratio at the
average anharmonicity. 

\begin{figure}
\includegraphics[width=0.5\textwidth]{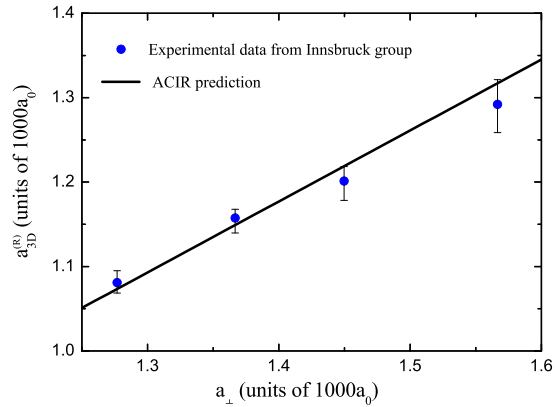}

\caption{(Color online) Predicted resonance positions of 2D ACIR resonances,
as a function of the transverse confinement parameter, $a_{\perp}$.
The experimental data points are taken from experiment \cite{Haller2010C},
where circles indicate resonances identified by absorption edges.
\label{fig:2DResonance_vs_a}}
\end{figure}

Finally, we plot the predicted variation of $a_{3D}^{(R)}$ with transverse
confinement parameter $a_{\perp}$, at a fixed value of $\omega\alpha$
corresponding to the Innsbruck experiments, in Fig.\ref{fig:2DResonance_vs_a}.
This data is also in excellent quantitative agreement with ACIR predicted
resonance positions.

In anharmonic trap experiments, we would generically expect both the
ACIR resonance at $a_{3D}>0$, and the internal HCIR resonance at
$a_{3D}<0$ to be observed. The Innsbruck experiments \cite{Haller2010C}
show clear evidence for the ACIR resonance, but not the internal HCIR
resonance. The Cambridge experiments \cite{Frohlich2010R} have shown
evidence for HCIR resonances, although the magnetic field scans were
not large enough to observe any ACIR resonances. The question of which
is observed will depend on the size of the anharmonic parameter, the
dynamics of the experiment, and the method of detection of the resonance.
It appears likely that the molecular loss technique is particularly
sensitive to COM resonances.

\section{conclusion\label{sec:conclusion}}

We have extended the theory of the harmonic confinement-induced resonance
to include the effects of anharmonic confinement, or ACIR, since previous
harmonic theories cannot explain phenomena observed in recent experiments.
In the presence of anharmonic perturbation of the confinement trap,
the COM motion of two atoms couples to the relative motion, and additional
resonances appear. We have calculated the energy of the resulting
new resonances up to first order in perturbation theory. The results
agree well with experiments, with both even and odd order multiple
resonances being found. These differences are not just small perturbations
on previous HCIR predictions, and show large qualitative differences
from the predictions for harmonic traps.

ACIR resonances due to COM excitations are always present in Feshbach
systems with any form of transverse confinement. The important issue
is that for harmonic traps the Kohn theorem prevents these resonances
from being coupled to incoming atoms in the two-body sector of the
COM ground state. However, experimental optical traps do have relatively
large anharmonic parameters, which allows the resonances due to COM
motion to be coupled to incoming scattering states. We find excellent
quantitative agreement between ACIR predictions and recent experimental
observations of confinement resonances, both in one and two dimensional
traps. \\

\begin{acknowledgments}
This work was supported by the ARC Discovery Projects DP0984637, DP0880404,
DP0984522 and NFRPC Grant No. 2011CB921502. One of us, P.D.D., wishes
to thank the Aspen Center for Physics for their generous hospitality.
We also wish to acknowledge useful discussions and communications
with C. Vale, H-C. Nagerl, R. Grimm, M. Kohl, Simon Sala, Philipp-Immanuel
Schneider, and Alejandro Saenz.
\end{acknowledgments}
\appendix*
\begin{widetext}

\section{Anharmonic Perturbation Theory}

We consider how the relative anharmonic energy of the virtual molecule
$\epsilon_{a}$ Eq.(\ref{eq:Harmonic threshold-1}), is calculated
from the anharmonic perturbation $H_{a}$, including the effects of
anharmonicity on both relative and COM motion. For space reasons,
we do not give all the overlap integrals, as the calculations are
similar in all cases. Instead, we focus on the most important cases.

First, note that the dimensionless change in the relative anharmonic
perturbation energy is: 
\begin{equation}
\epsilon_{a}=\frac{1}{\hbar\omega_{y}}\left[E_{M}^{(1)}-2E_{A}^{(1)}\right]\,,
\end{equation}
where $E_{A}^{(1)}$ is the perturbed atomic anharmonic energy given
in Eq (\ref{eq:One-atom-anharmonic}), and $E_{M}^{(1)}$ is the first
order perturbation theory correction to the energy of the virtual
molecule: 
\begin{eqnarray}
E_{M}^{(1)} & = & \left\langle \Psi_{N_{x}N_{y}}\psi_{b}\right|H_{a}\left|\psi_{b}\Psi_{N_{x}N_{y}}\right\rangle \,.
\end{eqnarray}
 Here, $\left|\Psi_{N_{x}N_{y}}\psi_{b}\right\rangle =\left|\Psi_{N_{x}N_{_{y}}}\right\rangle \left|\psi_{b}\right\rangle $
is the eigenstate of $H_{h}$. This is treated as the zero-order wave-function,
and includes both relative and COM motion. Next, we introduce $\left|\Psi_{N_{x}N_{_{y}}}\right\rangle $
as the unperturbed wave-function of the COM Hamiltonian, so that 
\begin{equation}
\left|\Psi_{N_{x}N_{y}}\right\rangle =\left|\phi_{N_{x}}\right\rangle \left|\phi_{N_{y}}\right\rangle \label{eq:1D 12}
\end{equation}
with the standard two-dimensional harmonic oscillator solution of:

\begin{equation}
\left|\phi_{N_{x,y}}\right\rangle =\frac{\exp\left(-2\xi_{x,y}^{2}\right)H_{N_{x,y}}\left(2\xi_{x,y}\right)}{\sqrt{\pi^{1/2}d_{x,y}2^{N_{x,y}-1}N_{x,y}!}}\label{eq:14}
\end{equation}
 where $\xi_{i}=\left(X/d_{x},Y/d_{y}\right)$. We now turn to the
task of evaluating $E_{M}^{(1)}$, the anharmonic correction to the
total molecular energy at a given unperturbed dimensionless energy
$\epsilon$.

\subsection{The relative wave-function $\left|\psi_{b}\right\rangle $ }

The specific form of $\left|\psi_{b}\right\rangle $ can be obtained
by directly solving the eigenproblem of $H_{h}^{rel}$ ,
\begin{equation}
\left[-\frac{\hbar^{2}}{2\mu}\nabla_{\mathbf{r}}^{2}+\frac{1}{2}\mu\omega_{x}^{2}x^{2}+\frac{1}{2}\mu\omega_{y}^{2}y^{2}+V\left(\mathbf{r}\right)\right]\left|\psi\right\rangle =E\left|\psi\right\rangle \label{eq:15}
\end{equation}
 The wave-function $\left|\psi\right\rangle $ can be written as,
\begin{eqnarray}
\left|\psi\right\rangle  & = & \psi_{0}\left(\mathbf{r}\right)-\int_{0}^{\infty}d\mathbf{r}^{\prime}G_{E}\left(\mathbf{r},\mathbf{r}^{\prime}\right)V\left(\mathbf{r}^{\prime}\right)\psi\left(\mathbf{r}^{\prime}\right)\nonumber \\
 & = & \psi_{0}\left(\mathbf{r}\right)-\mathcal{R}\cdot\frac{2\pi\hbar^{2}a_{3D}}{\mu}G_{E}\left(\mathbf{r},0\right)\label{eq:16}
\end{eqnarray}
 Where $\psi_{0}\left(\mathbf{r}\right)$ is the regular solution
of Eq.(\ref{eq:15}) and,
\begin{equation}
\mathcal{R}=\left[\frac{\partial}{\partial r}r\psi\left(\mathbf{r}\right)\right]_{r=0}\label{eq:17}
\end{equation}
 Here, we have use the pseudo-potential approximation. For a bound
state, the solution of Eq.(\ref{eq:15}) $\left|\psi\right\rangle $
should vanish as $r\rightarrow\infty$, hence $\psi_{0}\left(\mathbf{r}\right)=0$
. Then the constant $\mathcal{R}$ is only a normalization coefficient.
The Green's function $G_{E}\left(\mathbf{r},\mathbf{r}^{\prime}\right)$
is the solution of the following equation,
\begin{equation}
\left[-\frac{\hbar^{2}}{2\mu}\nabla_{\mathbf{r}}^{2}+\frac{1}{2}\mu\omega_{x}^{2}x^{2}+\frac{1}{2}\mu\omega_{y}^{2}y^{2}-E\right]G_{E}\left(\mathbf{r},\mathbf{r}^{\prime}\right)=\delta\left(\mathbf{r}-\mathbf{r}^{\prime}\right)\label{eq:18}
\end{equation}
 The form of the Green's function $G_{E}\left(\mathbf{r},0\right)$
for a bound state can be easily calculated,
\begin{eqnarray}
G_{\epsilon<0}\left(\mathbf{r},0\right) & = & \frac{\sqrt{\eta}I_{1}\left(\mathbf{r}\right)}{2\pi^{3/2}d^{3}\hbar\omega_{y}}\exp\left(-\frac{\eta x^{2}+y^{2}}{2d^{2}}\right)\label{eq:19}
\end{eqnarray}
where we introduce the integral representation:
\begin{equation}
I_{1}\left(\mathbf{r}\right)=\int_{0}^{\infty}dt\frac{\exp\left(\frac{\epsilon}{2}t-\frac{1}{t}\frac{z^{2}}{d^{2}}-\frac{e^{-\eta t}}{1-e^{-\eta t}}\frac{\eta x^{2}}{d^{2}}-\frac{e^{-t}}{1-e^{-t}}\frac{y^{2}}{d^{2}}\right)}{\sqrt{t\left(1-e^{-\eta t}\right)\left(1-e^{-t}\right)}}
\end{equation}
 Recalling that $\eta=\omega_{x}/\omega_{y}$ , and $E=\left(\epsilon+\eta/2+1/2\right)\hbar\omega_{y}$,
the bound-state wave-function is,
\begin{eqnarray}
\left|\psi_{b}\right\rangle  & = & \mathcal{N}\exp\left(-\frac{\eta x^{2}+y^{2}}{2d^{2}}\right)I\left(\mathbf{r}\right)\,,\label{eq:20}
\end{eqnarray}
where $\mathcal{N}$ is the normalization coefficient.

\subsection{1D COM overlap terms}

We can now calculate the overlap integrals which give the anharmonic
contributions to the quasi-1D and quasi-2D ACIR resonances. In order
to calculate the first order modification $E_{M}^{(1)}$ , we focus
on the $\left(N_{x},N_{y}\right)$ COM state. The relevant terms are:
$\left\langle \Psi_{N_{x}N_{y}}\right|X^{4}\left|\Psi_{N_{x}N_{y}}\right\rangle $
, $\left\langle \Psi_{N_{x}N_{y}}\right|Y^{4}\left|\Psi_{N_{x}N_{y}}\right\rangle $
, $\left\langle \Psi_{N_{x}N_{y}}\right|X^{2}\left|\Psi_{N_{x}N_{y}}\right\rangle $
, $\left\langle \Psi_{N_{x}N_{y}}\right|Y^{2}\left|\Psi_{N_{x}N_{y}}\right\rangle $. 

We define $D_{x,y}=\sqrt{\hbar/2m\omega_{x,y}}=d_{x,y}/2$ as the
transverse confinement parameter of an atom pair in the following
calculation. Hence:
\begin{eqnarray}
\left\langle \psi_{N_{x}N_{y}}\right|X^{4}\left|\psi_{N_{x}N_{y}}\right\rangle  & = & \frac{1}{\sqrt{\pi}D_{x}2^{N_{x}}N_{x}!}\int_{-\infty}^{\infty}X^{4}\exp\left(-\frac{X^{2}}{D_{x}^{2}}\right)\left[H_{N_{x}}\left(\frac{X}{D_{x}}\right)\right]^{2}dX\nonumber \\
 & = & \frac{1}{\sqrt{\pi}D_{x}2^{N_{x}}N_{x}!}\cdot D_{x}^{5}\sqrt{\pi}2^{N_{x}}N_{x}!\left[\left(N_{x}+\frac{1}{2}\right)\left(N_{x}+\frac{3}{2}\right)+\frac{1}{2}N_{x}\left(N_{x}-1\right)\right]\nonumber \\
 & = & \left[\left(N_{x}+\frac{1}{2}\right)\left(N_{x}+\frac{3}{2}\right)+\frac{1}{2}N_{x}\left(N_{x}-1\right)\right]D_{x}^{4}
\end{eqnarray}
 
\begin{eqnarray}
\left\langle \psi_{N_{x}N_{y}}\right|X^{2}\left|\psi_{N_{x}N_{y}}\right\rangle  & = & \frac{1}{\sqrt{\pi}D_{x}2^{N_{x}}N_{x}!}\int_{-\infty}^{\infty}X^{2}\exp\left(-\frac{X^{2}}{D_{x}^{2}}\right)\left[H_{N_{x}}\left(\frac{X}{D_{x}}\right)\right]^{2}dX\nonumber \\
 & = & \frac{1}{\sqrt{\pi}D_{x}2^{N_{x}}N_{x}!}\cdot D_{x}^{3}\sqrt{\pi}2^{N_{x}}N_{x}!\left(N_{x}+\frac{1}{2}\right)\nonumber \\
 & = & \left(N_{x}+\frac{1}{2}\right)D_{x}^{2}
\end{eqnarray}
 By using the exchange symmetry of $x$ and $y$ , we can easily obtain,
\begin{equation}
\left\langle \psi_{N_{x}N_{y}}\right|Y^{4}\left|\psi_{N_{x}N_{y}}\right\rangle =\left[\left(N_{y}+\frac{1}{2}\right)\left(N_{y}+\frac{3}{2}\right)+\frac{1}{2}N_{y}\left(N_{y}-1\right)\right]D_{y}^{4}
\end{equation}
\begin{equation}
\left\langle \psi_{N_{x}N_{y}}\right|Y^{2}\left|\psi_{N_{x}N_{y}}\right\rangle =\left(N_{y}+\frac{1}{2}\right)D_{y}^{2}
\end{equation}
 Here, we have used the formulas,
\begin{equation}
\int_{-\infty}^{\infty}dtt^{2}e^{-t^{2}}H_{n}^{2}\left(t\right)=\sqrt{\pi}2^{n}n!\left(n+\frac{1}{2}\right)
\end{equation}
\begin{equation}
\int_{-\infty}^{\infty}dtt^{4}e^{-t^{2}}H_{n}^{2}\left(t\right)=\sqrt{\pi}2^{n}n!\left[\left(n+\frac{1}{2}\right)\left(n+\frac{3}{2}\right)+\frac{1}{2}n\left(n-1\right)\right]
\end{equation}

\subsection{1D Internal motion overlap terms}

The relevant terms are now: $\left\langle \psi_{b}\right|x^{4}\left|\psi_{b}\right\rangle $
, $\left\langle \psi_{b}\right|y^{4}\left|\psi_{b}\right\rangle $,
$\left\langle \psi_{b}\right|x^{2}\left|\psi_{b}\right\rangle $ ,
$\left\langle \psi_{b}\right|y^{2}\left|\psi_{b}\right\rangle $.
For the bound state of the relative motion $\left|\psi_{b}\right\rangle $
, the main contribution of the integrals, \emph{e.g.}, $\left\langle \psi_{b}\right|x^{4}\left|\psi_{b}\right\rangle $
\emph{etc.}, comes from the regime around the origin. We note that
the energy parameter $\epsilon$ is negative. The form of $\left|\psi_{b}\right\rangle $
near the origin $r=0$ can be easily obtained from Eq.(\ref{eq:20}),
\begin{eqnarray}
\left|\psi_{b}\right\rangle  & \approx & \left(\frac{-\epsilon}{2}\right)^{1/4}\frac{1}{\sqrt{\pi d_{y}}}\cdot\frac{\exp\left(-\sqrt{-2\epsilon}r/d_{y}\right)}{r}
\end{eqnarray}
 which has been normalized. Then,
\begin{eqnarray}
\left\langle \psi_{b}\right|x^{2}\left|\psi_{b}\right\rangle  & = & \frac{1}{\pi d_{y}}\sqrt{\frac{-\epsilon}{2}}\int_{0}^{2\pi}\int_{0}^{\pi}\int_{0}^{\infty}r^{2}\sin^{3}\theta\cos^{2}\phi\exp\left(-\frac{2\sqrt{-2\epsilon}r}{d_{y}}\right)drd\theta d\phi\nonumber \\
 & = & -\frac{d_{y}^{2}}{12\epsilon}
\end{eqnarray}
\begin{eqnarray}
\left\langle \psi_{b}\right|x^{4}\left|\psi_{b}\right\rangle  & = & \frac{1}{\pi d_{y}}\sqrt{\frac{-\epsilon}{2}}\int_{0}^{2\pi}\int_{0}^{\pi}\int_{0}^{\infty}r^{4}\sin^{4}\theta\cos^{4}\phi\frac{\exp\left(-2\sqrt{-2\epsilon}r/d_{y}\right)}{r^{2}}r^{2}\sin\theta drd\theta d\phi\nonumber \\
 & = & \frac{3d_{y}^{4}}{40\epsilon^{2}}
\end{eqnarray}
 
\begin{eqnarray}
\left\langle \psi_{b}\right|y^{2}\left|\psi_{b}\right\rangle  & = & \frac{1}{\pi d_{y}}\sqrt{\frac{-\epsilon}{2}}\int_{0}^{2\pi}\int_{0}^{\pi}\int_{0}^{\infty}r^{2}\sin^{2}\theta\sin^{2}\phi\frac{\exp\left(-2\sqrt{-2\epsilon}r/d_{y}\right)}{r^{2}}r^{2}\sin\theta drd\theta d\phi\nonumber \\
 & = & -\frac{d_{y}^{2}}{12\epsilon}
\end{eqnarray}
\begin{eqnarray}
\left\langle \psi_{b}\right|y^{4}\left|\psi_{b}\right\rangle  & = & \frac{1}{\pi d_{y}}\sqrt{\frac{-\epsilon}{2}}\int_{0}^{2\pi}\int_{0}^{\pi}\int_{0}^{\infty}r^{4}\sin^{4}\theta\sin^{4}\phi\frac{\exp\left(-2\sqrt{-2\epsilon}r/d_{y}\right)}{r^{2}}r^{2}\sin\theta drd\theta d\phi\nonumber \\
 & = & \frac{3d_{y}^{4}}{40\epsilon^{2}}
\end{eqnarray}

\subsection{The first order modification of the energy $E_{M}^{(1)}$}

Using the overlap integral results given above, we find that:
\begin{equation}
\left\langle \psi_{N_{x}N_{y}}\psi_{b}\right|H_{a}^{x}\left|\psi_{N_{x}N_{y}}\psi_{b}\right\rangle =\frac{\alpha_{x}\hbar\omega_{x}}{32}\left[6N_{x}\left(N_{x}+1\right)+3-\frac{\eta\left(2N_{x}+1\right)}{\epsilon}+\frac{3\eta^{2}}{10\epsilon^{2}}\right]\label{eq:x-anharmonic energy}
\end{equation}
\begin{equation}
\left\langle \psi_{N_{x}N_{y}}\psi_{b}\right|H_{a}^{y}\left|\psi_{N_{x}N_{y}}\psi_{b}\right\rangle =\frac{\alpha_{y}\hbar\omega_{y}}{32}\left[6N_{y}\left(N_{y}+1\right)+3-\frac{\left(2N_{y}+1\right)}{\epsilon}+\frac{3}{10\epsilon^{2}}\right]\label{eq:y-anharmonic energy}
\end{equation}

Combining this with the atomic energy correction gives the overall
result for an equal wavelength 1D optical lattice, in which $\alpha_{x}\omega_{x}=\alpha_{y}\omega_{y}$
:
\begin{equation}
\epsilon_{a}^{(1D)}=\frac{\alpha}{32}\left[6(N_{y}^{2}+N_{y}+N_{x}^{2}+N_{x})-18-\frac{\eta\left(2N_{x}+1\right)+\left(2N_{y}+1\right)}{\epsilon}+\frac{3\left(\eta^{2}+1\right)}{10\epsilon^{2}}\right]
\end{equation}

Following similar procedures, from Eq \ref{eq:y-anharmonic energy},
the anharmonic correction in the corresponding 2D case leads to: 

\begin{equation}
\epsilon_{a}^{(2D)}=\frac{\alpha}{32}\left[6N\left(N+1\right)-9-\frac{\left(2N+1\right)}{\epsilon}+\frac{3}{10\epsilon^{2}}\right]
\end{equation}

.

\end{widetext}

\end{document}